# Strong light/matter coupling for SERS

Giuseppina Simone

For a wide range of health applications, label-free sensing is essential, and micro-photonics offers innovative technical solutions to pursue important objectives. It has been shown that the strong light/matter coupling formed between a cavity and molecules through light excitation enhances biological and clinical applications by providing deep insights into molecular analysis.
The multilayer cavity's proposed architecture is meant to promote a robust interaction between light and matter. The top layer was made of silver Ag multi-shape features that, after being synthesized and characterized, were immobilized on the surface, while a layer of indium tin oxide ITO was flipped to produce two distinct layouts. An investigation of the mode behavior was conducted to describe the two layouts; experimental and numerical results point to a strong light/matter interaction by the ITO/SiO2/spacer/Ag multi-shape feature design. For the characterization of the light/matter coupling, a fluorophore was deposited on the surface; the anticrossing energy was then examined by integrating the experimental data with a model of three mechanical oscillators. To show the system's sensitivity, the analysis was carried out again using bovine serum albumin (BSA) protein. The protein is water-soluble and exhibits an infrared absorption band (amide I), while also being active in the Raman region. Besides, it may consistently bind to Ag multi-shape features. The excellent sensitivity was demonstrated, enabling the use of image analysis to capture the surface-enhanced Raman scattering of the BSA. In conclusion, the suggested sensing approach brings up fresh possibilities for highly sensitive biomolecule detection techniques and encourages results when used as a fundamental sensing technique to investigate molecular patterns.

## 1. Introduction

Highly sensitive detection of molecules is crucial for several applications, such as diagnosis, drug discovery, and security screening. In the last decades, optical detection has garnered a lot of attention due to the high-$Q$ resonances that have been attained.[1,2] However, to study the atom-cavity interaction regime and have a substantial impact on achieving high sensitivity with limits down to the molecular scale, ultra-high $Q$ resonances are required.[3–5]

The sensing mechanism of multilayer cavities critically depends on the quality factor $Q$ in the sense that the coupling between molecules and a high $Q$ cavity perturbs the hybrid cavity mode by leading to the shift of the resonance and altering the cavity transmission, and the capability to detect the shift measures the sensitivity of the system. The most recent result reports that a resonance shift of about $10^{-10}$ with a cavity having $Q = 10^6$ in an aqueous environment can be achieved.[6] This milestone represents an important result in molecular analysis, but it is still far from the sensitivity necessary for detecting a single binding event. Thus, in molecular optomechanics, control of optical and mechanical resonator behavior is critical for enhancing sensitivity.

Fabry-Perot cavities are accounted for as optical resonators that have been successfully conceptualized for applications in quantum physics, as well as chemical sensing and metrology.[7,8] The fundamental layout of the optical resonator consists of two mirrors as a reflective element, separated by a transparent layer, the cavity. The mirrors establish symmetry, enabling the waves to travel within the cavity; besides, because the mirrors have a modulus close to zero, the waves behave like standing ones. The Fabry-Perot cavity displays high-quality factor $Q$ and small modal volume $V/V_0$ and for this reason, they play a decisive role in quantum electrodynamics investigation and light/matter interaction.[9,10] The high $Q$-factor derives from a linewidth that is narrower than the vibrational frequency of the molecular species and imposes the crucial hierarchy in a resolved side-band regime where the mechanical frequency exceeds the optical linewidth as a fundamental condition.[11] An important family of Fabry-Perot systems is constituted by the metal/dielectric/metal stack layer cavity, which is frequently used in cutting-edge plasmonics research. These systems display resonant behavior when the cavity thickness is comparable to the wavelength, while to optimize the quality factor, the metallic mirror thickness must be kept high. This approach prevents the majority of the pump radiation from reaching atoms inside the cavity, hindering the possibility of using a metal/dielectric/metal resonator in applications in which the inclusion of a photo-sensible element is required.

An aspect to consider is that the higher sensitivity related to a high $Q$-factor represents a limitation because of the losses associated with the resonance of the system. Under resonance, the linewidth becomes much larger than the frequency, with a consequently conspicuous decay rate.[12–14] Therefore, the losses limit the performance of the figure-of-merit of the resonant system and negatively impact the Q-factor. To reduce the linewidth of the plasmonic resonance, several strategies based on plasmonic hybridization have been studied.[15–17] Hybrid plasmonic systems including dielectric and metal layers combined with nanometric features allow for design tuning and flexibility, which results in a variety of resonance line shapes and cavity modes.[18,19] Besides, the hybridization between the SPPs and the cavity requires a higher level of complexity to be understood;[20] for example, the optical antenna offers excellent confinement of the electric field for large coupling to the molecular dipole,[16,17] at the price of a significant reduction of the effective detection area.

In order to examine the characteristics of sensing a multilayer cavity, here is a transparent hybrid system made up of a cavity with silver (Ag) multi-shape features that serve as the metallic mirror as well as plasmonic nanometric features for determining the system's plasmonic characteristics. As a second mirror, indium tin oxide (ITO) is employed, which optimizes the cavity design by alternating high-quality optical mirrors with good reflectivity with the dielectric layer. The mirror heights are designed to be significantly larger than the surface plasmon polariton penetration depth in the air so that losses are reduced.

The investigation gains value when biomolecules are coupled to the hybrid system. Indeed, biomolecule coupling is more difficult since it calls for matching the mechanical energy provided by molecular vibration. Indeed, although the strong vibrational coupling of biomolecules has been established, there are still some difficulties to be solved due to the lower extinction coefficient of vibrational transitions than that of electronic transitions. This suggests that either high sample concentrations or very highly sensitive optical oscillators that can detect fluctuations are needed. To date, the bovine serum albumin protein, BSA, has been tested, and a strong coupling has been demonstrated for this complex system.[21]

In conclusion, there is an outstanding advantage related to the success in achieving the coupling of light/matter in the present configuration. Indeed, the interaction takes place between the single plasmon mode of the cavity and the molecular mode. The cavity and the molecules work as mechanical resonators with significantly different typical frequencies (several MHz for the cavity and a few tens of GHz for the molecules), on which the capability of sensitive probing and interaction of the system with the surrounding environment depends. Furthermore, in the form of SPPs, the light hybridizes the cavity, couples the molecule, and triggers the plasmon-exciton coupling, displaying the molecular vibrational modes and disclosing a region rich in molecular Raman transitions.[22–24]

Based on these considerations, cavity/resonator hybridization is expected to enhance the sensitivity of molecular analysis.

## 2. Results
### 2.1 Multi-layer cavity plasmonic characterization

To prepare the plasmonic samples, the functional mode (3-mercaptopropyl) trimethoxy-silane, which served as the functional molecule, was used to create a chemical bridge between the Ag features and the surface of an ITO/$SiO_2$ substrate (**Figure 1a**). In this investigation, ITO was chosen for its transparency in the visible spectrum, and silver (Ag) was chosen for its well-known reflectance behavior, and it consisted of multi-shape features (**Figure 1b**). The (3-mercaptopropyl) trimethoxy-silane (designated as "spacer" in the image) was used to coat the $SiO_2$, Ag multi-shape feature concentration per area was lower than $10^4$ particles cm$^{-2}$.

For the plasmonic measurements, a monochromatic *p*-polarized light source was used to illuminate the structure from the side opposite to the Ag features, and the impinging angle was adjusted. The layout has two major modes that can be distinguished independently (**Figure 2a**); these are labeled m1 and m2 and occurred at $\vartheta_1$=44.9 degrees and at $\vartheta_2$=48.8 degrees. The result contrasts with the reflectance spectrum of the Ag multi-shape features adorning the surface of a standard $SiO_2$ cover glass (dashed curve in Figure with a dip at $\vartheta_{res}$=39.7 degrees), which is characterized only by a single mode that can be attributed to Ag. The modes were analyzed at the two excitation configurations relative to the s- and p-polarization. At an impinging angle $\vartheta_2$=44.9 degrees (panel (i)), the mode m2 is excited and visible with the same intensity at both the p- and s- polarization, while the mode m1 requires illumination with the p-polarization to show a symmetrical distribution along the x- and y-axes. At an impinging angle $\vartheta_1$=48.8 deg (panel (ii)), the modes display a higher intensity than the previous case; moreover, the maximal intensity is highly confined around the borders of the Ag feature, and both M1 and M2 are more distinct at p-polarization.

Overall, the findings in relation to the three scenarios show that the surface plasmons' phase symmetry defines how the cavity is excited. At the normal incidence, the electric field at the opposite edges of the cavity has the same phase along the direction perpendicular to the polarization and opposite phase along the polarization. From this trend, it can be inferred that the m1 mode prospect to have a relevant impact on the system's sensitivity; indeed, the configuration relative to the illumination cannot always make up for the phase difference of the gap plasmons produced at the opposite edges, and this challenges the visualization of the shallow cavity modes in the far-field spectra. The analysis of the dispersion displays the energy minima at the angles that have been determined by the experimental results (**dashed lines in Figure 2c**). In contrast to what a typical Fabry-Perot cavity would suggest and what the behavior of the mode m1 supports, the mode m2 redshifted when the angle increased.

A hysteresis diagram is produced by the relationship between the amplitude and phase of the reflectance signal (**Figure 2d**). At the origin of the XY axis, the two branches of the curve relating to the sample A shows a crossover that results in a double hysteresis loop due to high electric energy storage (area included between the dashed axes and the lobe of the loop) that constrasts a reduced energy loss (area of the lobe). Because the energy density has an important role on the plasmonic behavior, it is expected to support the capability to form a strong light/matter coupling. A conclusion that can now be confirmed is that the various layer structures exhibit varying behavior that is not just based on the naive material's dielectric constant but instead has a deeper impact on the system's reflectivity, which is tied to the stacking order.[25]

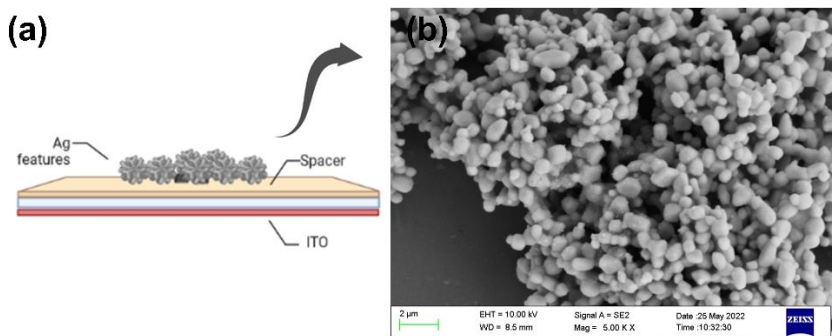

**Figure 1.** Plasmonic cavity. (a) layout. (b) Scanning electron image of the Ag multi-shape features used a top layer.

The electric and magnetic field components normalized to the amplitude of the incoming light are described by the maps in **Figure 2e** and **Figure 2f**, which focus on the Ag surface. Regardless of the mode (m1 or m2), the zeta component of the electric field $E_z$ exhibits a high intensity around the borders of the Ag feature; nevertheless, the radiation is stronger under the m1 mode settings than in the m2 mode. For the two modes, the magnetic field oscillates in the vertical z-direction; however, in the case of mode m2, the magnetic field is entirely contained within the multilayer, and in the case of mode m1, it is also strongly developed around the Ag feature. Additionally, for mode m1, the intensity increases at the spacer between the Ag multi-shape features and the $SiO_2$ interface, resulting in a greater current distribution than under the conditions of the m2 mode. With reference to the $H_y$ component of the magnetic field, because the energy behavior can be modeled by a classical harmonic oscillator, the enhancement field that is recorded in the spacer region must be the result of constructive in-phase interference of the magnetic field between the Ag mode and cavity mode. In turn, the lower intensity may be due to the magnetic field's orientation changing in the spacer region, which results in the field enhancement factor weakening.

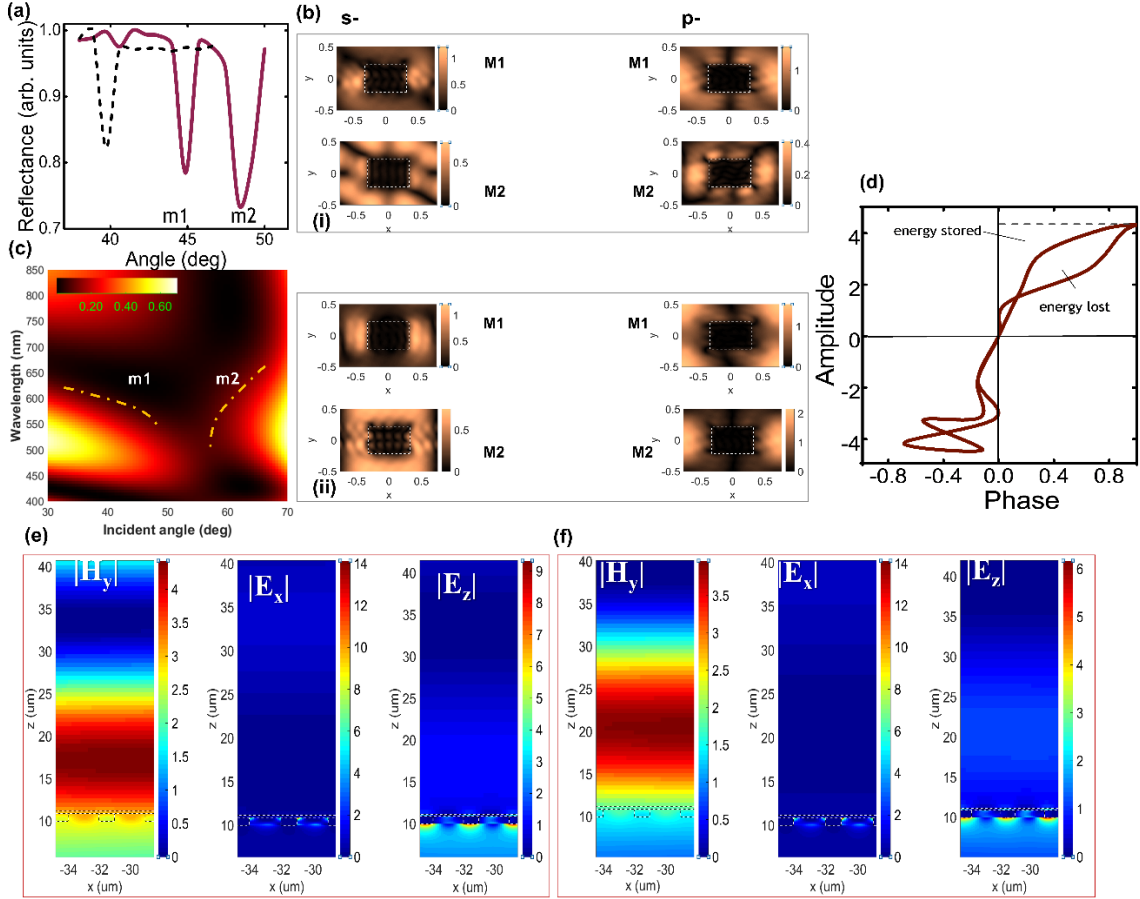

the test biomolecule because it can reliably attach to Ag multi-shape features and exhibits an infrared absorption band (amide I) that is simultaneously active in the Raman region. For comparison, Rhodamine was tested as well (Figure 3a); the oscillation of the cavity causes a clear shift that may be related to the various molecular weights. For example, BSA, which has a molecular weight of 66.5 kDa, shifts by around 35 kHz, whereas Rhodamine G6, which has a molecular weight of 4.8 kDa, shifts by about 20 kHz.

**Figure 2.** Plasmonic behavior. (a) *p*-polarized reflectance measured at $\lambda_{exc}$ =650 nm. (b) Mode analysis (i) at $\vartheta$=44.9 degrees, (ii) at $\vartheta$=48.8 degrees. (c) Reflectance dispersion. (d) Hysteresis of the amplitude/phase diagram. (e) electromagnetic field, at $\vartheta$=44.9 degrees. (f) electromagnetic field, at $\vartheta$=48.8 degrees.

**2.3 Light-matter strong coupling with a protein: spontaneous emission enhancement and SERS**

In light of the earlier findings and their analogy with the behavior of a Fabry-Perot cavity, the analysis of the strong coupling in relation to layout A is the focus of the investigation that follows. The architecture of the metal/dielectric platform is an ideal layout for studying the interaction between the excitons of the molecules and the polaritons of the cavity because the design allows the excitons to be accessible to the light beam. For demonstrating the validity of this hypothesis, the coupling between the cavity/protein system was studied. The coupling with the biomolecules is more complicated than with emitters (e.g. dyes, quantum dots) because the extinction coefficient of vibrational transitions is lower than that of electronic transitions. Therefore, if, on one side, the strong vibrational coupling of biomolecules has been demonstrated,[26] there are still some challenges that must be overcome that are related to the smaller extinction coefficient of vibrational transitions compared to those of electronic transitions. This observation brings us to the conclusion that to sense biomolecules using the strong coupling principle, either large sample concentrations are needed or else extremely sensitive optical oscillators are needed.

The optomechanical spectrum was measured to evaluate molecular sensitivity, bovine serum albumin (BSA) was chosen as

To date, the protein solution has been dripped onto the surface of the A cavity for measuring the reflectance, which is dramatically reduced in intensity as well as shifted to higher angles (Figure 3b). When studying the dispersion of the system, the lower branch shifts and increases intensity as the angle of incidence increases, while the upper branch initially loses intensity and then disperses to higher energy (Figure 3c). An anticrossing is observed when the energy of the cavity, which has been widely characterized above, matches the energy of the protein, $E_m \sim 2.54 - 2.95\ eV$,[27] at an angle $\vartheta = 49\ degrees$ (Figure 3d). The anticrossing energy of the oscillator is $\hbar\Omega = 1.13$ eV, while a single exciton model can be used to verify the strong coupling. The transmission spectrum of the BSA/cavity features two peaks symmetrically spaced out around the dip at $\vartheta_i = 43$ degrees (Figure 3e) that can be understood as the spectral hallmarks of the two new hybrid polaritonic states.[28] Moreover, the comparison of the cavity amplitude and phase emission with and without the protein provides further information to support these findings (Figure 3f). It demonstrates how the BSA shifts the phase while simultaneously amplifying its module.[29] A comparison between the modes of the cavity coupled with Rhodamine and with the BSA is shown in Figure 3g. The cavity, including the Rhodamine, displays an electric field that is highly confined near the edge of the Ag feature, and it extends in the spacer as well as under the feature (left, panel (i)); in turn, the magnetic field band stretches along the y-axis direction (right, panel (i)) and has field hotspots at the edges of the Ag feature. Such electromagnetic field distributions correspond to the excitation of SPPs; both components of the field display localized hotspots at the edges of the Ag, which feature significant evidence of an intense coupling between the excitons and the polaritons.

As it has been extensively demonstrated, surface-enhanced Raman scattering significantly relies on electromagnetic enhancement that influences the pattern of the emitted signal in the presence of matter and molecules. Recently, Raman signals have been demonstrated to depend on the enhancement of the optomechanical coupling between the localized surface plasmon resonance and the vibrational mode of the molecule, because the strong light/matter interactions are caused by the hybridization of the electronic or vibrational transition with the modes of an optical cavity. In principle, strongly coupled cavity/vibrating molecules are modeled as a mechanical oscillator with absorbing infrared vibrational modes, which can be probed by a visible laser through their Raman scattering. The strong coupling allows the coherent conversion of signals between optical and mechanical domains, while the confinement of the signal in a tight cavity enhances the signal from however, the position of the excitons and polaritons is relatively close and the coupling is strong, as exhibited by the intensity of the field enhancement. The maximum electric and magnetic fields are enhanced to be about 600 and 4000 times the incident field, respectively. An analogous behavior is recorded when the BSA is taken into consideration; the maximum electric (left, panel (ii)) and magnetic (right, panel (ii)) fields are enhanced to be about 800 and 2500 times the incident field, respectively. Besides, the bonds and increases the efficiency of the system. These mechanical oscillators' strength depends on the distinct frequency ranges that the cavities and molecules exhibit (several MHz for the cavity and a few tens of GHz for the molecules). The system's capacity for probing and its interactions with its environment are both impacted by the disparity between the two frequency ranges. The amplification of the transmitted signal's positive impact on molecular analysis and sensing the BSA was quantified through the measurement of the Raman signal. A CCD camera was used to record the emitted signal, and the data were extrapolated from the pictures using the technique outlined in ref.[30] for producing a Raman spectrogram of the BSA. The heat map of emission intensity from consecutive measurements over time is shown in Figure 4a, while the top curve in Figure 4b displays the average surface-enhanced Raman scattering spectrum from 500 frames that were used for generating the heat map introduced above. It displays three consistent bands centered at 955 cm$^{-1}$, 1159 cm$^{-1}$, and 1550 cm$^{-1}$ that correspond to the stretching vibration of the C-N bond in the phenylalanine residues (Phe), the C-N stretching vibration of phenylalanine and acid amides, and the tyrosine-phenylalanine residues, and confirm a coherent result with the BSA SERS trace achieved by a standard instrument (Figure 4b, bottom). A more accurate comparison between the on-resonance cavity and a standard SERS spectrum underlines a superior sensitivity in this first scenario, prospecting the possibility of capturing a pattern of the spectrum richer in Raman transitions.

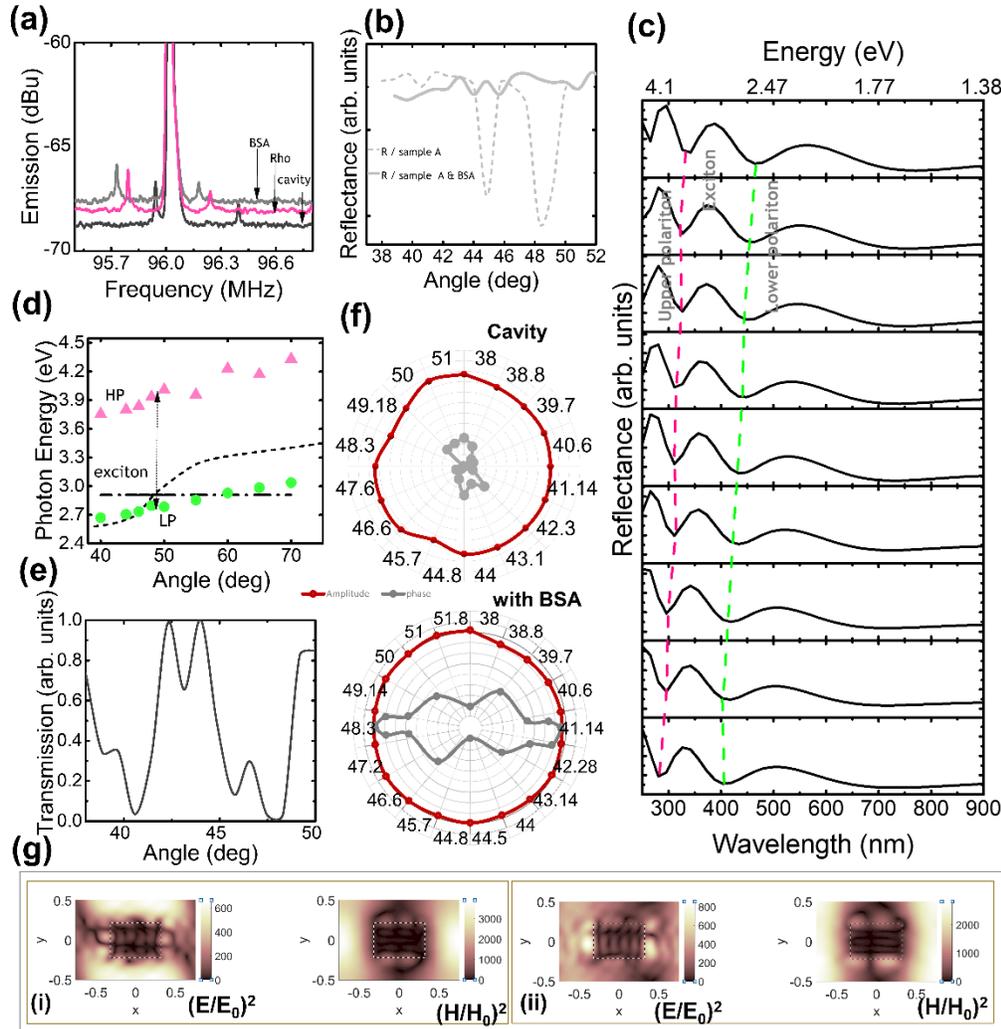

**Figure 3**. BSA/cavity. (a) Optomechanical spectra. (b) $p$-polarized reflectance according to the angle at $\lambda_{exc} = 650$ nm. (c) Experimental $p$-polarized reflectance at different angles from 40 to 78 degrees, from top to bottom $\Delta\vartheta = 4$ degrees. The mode of hybridization and generation of upper polariton, middle polariton, and lower polariton branches are displayed in the figure. (d) Experimental (dots) angular dispersion; the dashed lines represent the exciton and the cavity. (e) Emission of the system at $\lambda_{exc} = 630$ nm. (f) Polar diagram according to the impinging angle of the phase and amplitude of the emitted signal without cavity and with the BSA. (g) Mode analysis. (i) Rhodamine/cavity, (ii) BSA/cavity.

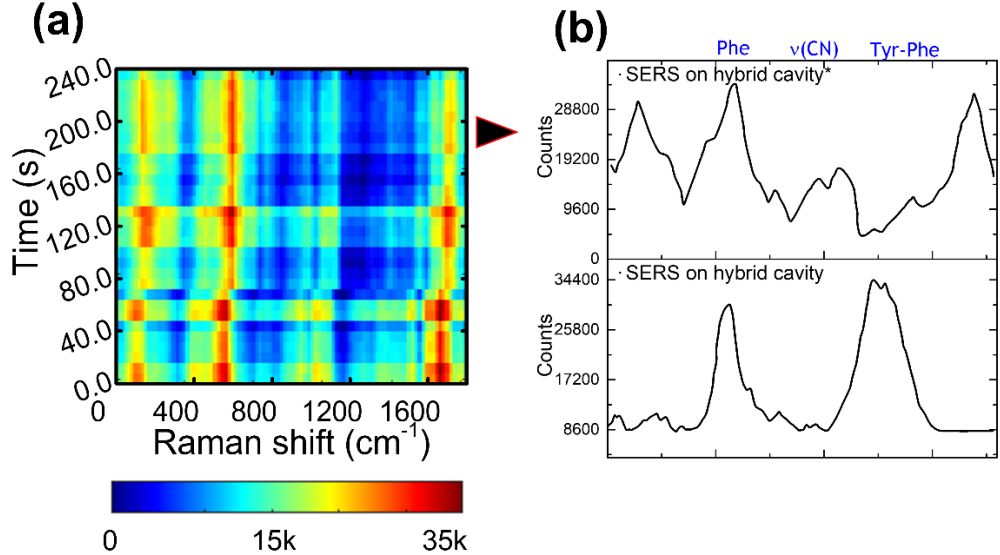

**Figure 4.** (a) Time-resolved spectroscopy of an imaged BSA/cavity system. (b) Average spectrum from panel (a) (top curve*), standard SERS (bottom) of BSA/cavity.

### 3. Conclusions

It has been demonstrated that a multilayer cavity can be used to explore molecular optomechanics. To investigate the coupling and gauge the strength, the BSA was tested. it was proved that the enhancement of the electromagnetic field intensity related to the strong coupling must influence the pattern of the emitted signal. To quantify the beneficial influence of the enhancement of emitted signal on the molecular analysis and on sensing the BSA, a measurement of the Raman signal was done. The results underlined a superior sensitivity of the method proposed here, prospecting the possibility of capturing a pattern of the spectrum richer in Raman transitions.

### 4. Experimental

*Synthesis of the Ag multi-shape features and chemical characterization.* The (3-mercaptopropyl) trimethoxy-silane (Sigma Aldrich, China) was used to graft the Ag multi-shape features to create the optical resonators. It was deposited by chemical vapor deposition (5 μl at 70 °C and 14.7 psia for 30 minutes) in a homemade furnace including a vacuum system and a temperature and pressure control. Cover slides with an ITO-coated side were utilized as the substrate, and the (3-mercaptopropyl) trimethoxy-silane was placed on the $SiO_2$. The substrates were cleaned with a mild stream of toluene or methanol and then dried in a gentle airflow. The pure Ag features solution (100 L, 0.1 mg $mL^{-1}$) was poured on top of the thiol and allowed to dry for 24 hours at room temperature. The substrate was treated with ultrasonication for 30 min in water to remove the multi-shape features which did not graft on the substrate.
*Optical analysis.* The detection configuration is composed of two moduli: SPR detection system, in reflection, and scattering detection system, in transmission. A detailed description of the experimental setup architecture and the schematic diagram was presented elsewhere. [31–34] Because the two detection systems are located on different sides of the prism, the layout allowed for a limited impact of the SPR on the SERS data. The SPR system was located on the rear of the prism (Thorlabs Inc.). The light source was a 0.7 mm dot laser (Shenzen Futhe Tech. Co. Ltd, power 5mW and multiple wavelengths), which was collimated (Thorlabs Inc. RC12FC-P01) and polarized by a double Glan-Taylor Calcite Polarizer, while the noise was lowered with a K-chopper. Before being collected by a silicon photodiode and delivered to an oscilloscope, the reflected beam was collimated and filtered (Thorlabs, Glan Thomson polarizer BBAR: 650-1050 nm, CA: 10 mm). The front portion of the prism held the Raman module. A silicon photodiode (FDS 1010 Thorlabs, λ=340-1100 nm, peak $\lambda_0$ = 960 nm responsivity $\Re(\lambda_0)$ 0.62 A $W^{-1}$) was used to gather the light, and a ceramic disk capacitor was used to cut down on noise. A spectrum analyzer is attached to the photodiode (PicoTech, series 5000). A Charge Coupled Device CCD was employed to detect the emitted signal for Raman's analysis. The theoretical simulation's code was created using the MATLAB programming language (MATLAB R2021a). For all experiments, an optical fluid (n= 1.515, Shanghai Specimen and Model Factory, China) allowed matching the refractive index of the device with an optical prism (Thorlabs; UV Fused Silica 25 mm right angle and n=1.466 at λ = 630 nm).

*The coupling model*
The dispersion of the hybridized mode was achieved by solving the eigenvalue problem of the three coupled oscillator model represented by the Hamiltonian of 2×2 matrix:

$$\begin{pmatrix} E_{spp} & G \\ G & E_m \end{pmatrix} \begin{pmatrix} \gamma_{spp} \\ \gamma_m \end{pmatrix} = E \begin{pmatrix} \gamma_{spp} \\ \gamma_m \end{pmatrix}$$

Here, Espp and Em are the energies of the uncoupled plasmon and monomeric exciton energy and E represents the eigenvalues relative to the energies of the hybrid. The eigenvalues of the Hamiltonian equation (1) has the following expression $E_\pm = \frac{1}{2}\left(E_{spp} + E_m + -\frac{i}{2}(\gamma_{spp} + \gamma_m)\right) \pm \frac{1}{2}\sqrt{4G^2 + \left(E_{spp} - E_m - \frac{i}{2}(\gamma_{spp} - \gamma_m)\right)^2}$ and accounts for the detuning energy and provides the Rabi-splitting model $\hbar\Omega = \sqrt{4G^2 - \frac{(\gamma_{spp} - \gamma_m)}{4}}$ when $E_{spp} = E_m$ ( G the coupling strength).

*Numerical analysis*

The numerical analysis was developed in Matlab environment. The script implements a frequency-domain modal method (known as the Rigorous Coupled wave Analysis/RCWA) in agreement with the model presented in ref [35]. It calculates the diffracted amplitudes and diffraction efficiency of structures with finite sizes made up of stacks of layers. The RCWA relies on the computation of the eigenmodes in all layers of the structure on a Fourier plane-wave basis [36] and a scattering matrix approach to recursively relate the mode amplitudes in the different layers.

The model consisted of the following layers, with the $p$-polarized light incident from the prism side: NBK7 substrate/ITO 1200 nm/$SiO_2$ 0.55 mm/1 μm Ag multi-shape features. The refractive index for each layer was taken from the dedicated source: Ag and ITO from Palik,[37] the NBK7 glass from Schott. In the simulation, the incident light was a normal TM-polarized plane wave and the experimental parameters were used for calibrating the model.